\documentstyle[aps,prl,epsf,floats]{revtex}

\newcommand{\la}[1]{\label{#1}}
\newcommand{\beq}{\begin{equation}}
\newcommand{\eeq}{\end{equation}}
\newcommand{\bea}{\begin{eqnarray}}
\newcommand{\eea}{\end{eqnarray}}
\newcommand{\be}{\begin{equation}}
\newcommand{\ee}{\end{equation}}
\newcommand{\ba}{\begin{eqnarray}}
\newcommand{\ea}{\end{eqnarray}}
\newcommand{\bi}{\begin{itemize}}
\newcommand{\ei}{\end{itemize}}
\newcommand{\rmi}[1]{{\mbox{\scriptsize #1}}}
\newcommand{\fig}{Fig.~}
\newcommand{\eq}{Eq.~}
\newcommand{\eqs}{Eqs.~}
\newcommand{\nr}[1]{(\ref{#1})}
\newcommand{\tr}{{\rm Tr\,}}
\newcommand{\nn}{\nonumber \\}
\newcommand{\fr}[2]{{\frac{#1}{#2}}}
\newcommand{\msbar}{\overline{\mbox{\rm MS}}}
\newcommand{\lambdamsbar}{\Lambda_{\overline{\rm MS}}}

\newcommand{\bmu}{\bar{\mu}}

\def\mb#1         {\mbox{\boldmath $#1$}}

\def\lsi{\raise0.3ex\hbox{$<$\kern-0.75em\raise-1.1ex\hbox{$\sim$}}}
\def\gsi{\raise0.3ex\hbox{$>$\kern-0.75em\raise-1.1ex\hbox{$\sim$}}}
\newcommand{\lsim}{\mathop{\lsi}}
\newcommand{\gsim}{\mathop{\gsi}}

\begin{document}

\twocolumn[\hsize\textwidth\columnwidth\hsize\csname
@twocolumnfalse\endcsname
%\draft

\title{How to resum long-distance contributions to the QCD pressure?}
\author{K. Kajantie$^{\rm a}$, M. Laine$^{\rm b,a}$, 
K. Rummukainen$^{\rm c,d}$, Y. Schr\"oder$^{\rm a}$}
\address{$^{\rm a}$Department of Physics,
P.O.Box 9, FIN-00014 University of Helsinki, Finland}
\address{$^{\rm b}$Theory Division, CERN, CH-1211 Geneva 23,
Switzerland}
\address{$^{\rm c}$NORDITA, Blegdamsvej 17,
DK-2100 Copenhagen \O, Denmark}
\address{$^{\rm d}$%
Helsinki Institute of Physics, 
P.O.Box 9, FIN-00014 University of Helsinki, Finland}
\date{September, 2000} %\today}
\maketitle

%\vspace*{-4.0cm}
%\noindent
%\hfill \mbox{CERN-TH/2000-198, NORDITA-2000/62HE, hep-ph/0007109}
%\vspace*{3.8cm}

\begin{abstract}\noindent
The strict coupling constant expansion for the free energy of hot QCD plasma 
shows bad convergence at all reasonable temperatures, and does not agree 
well with its 4d lattice determination. This has recently
lead to various refined resummations, whereby the agreement
with the lattice result should improve, at the cost of a loss of 
a formal agreement with the coupling constant expansion and 
particularly with its large infrared sensitive ``long-distance'' 
contributions. We show here how to resum the dominant long-distance effects 
by using a 3d effective field theory, and determine their magnitude 
by simple lattice Monte Carlo simulations.  \\

CERN-TH/2000-198, NORDITA-2000/62HE \hfill hep-ph/0007109

\vspace*{-0.6cm}
\end{abstract}

%PACS numbers: 
%12.38.Mh, %        Quark-gluon plasma
%11.10.Wx, %        Finite temperature field theory
%12.38.Gc, %        Lattice QCD calculations
%11.10.Kk, %        Field theory in dimensions other than four.
%\\
%Keywords:
%QCD,
%finite temperature,
%dimensional reduction,
%quark gluon plasma,
\vskip1.5pc]

\noindent
{\bf Introduction.}
At temperatures above 200 MeV, the properties 
of matter described by the laws of QCD are expected to change. 
The system should look more like a collection of free
quarks and gluons than a collection of their bound states, such as 
mesons. It is a challenge to find observables which would clearly
manifest this change, and hopefully also be directly or
indirectly measurable in heavy ion collision experiments. 

{}From the theoretical point of view, one of the simplest 
observables witnessing the change is the free energy of the 
plasma, or its pressure \cite{boyd}.
Indeed, according to the Stefan-Boltzmann law, 
the value of the free energy counts the number of light elementary 
excitations in the plasma, be they quarks and gluons, or mesons. 

The reality is somewhat more complicated.
Interactions change the Stefan-Boltzmann law, so that pressure
is no longer proportional to the number of degrees of freedom.
And in fact, interactions are strong. An explicit computation
of the free energy to order ${\cal O}(g^5T^4)$~\cite{az,zk,bn}
shows that there are large corrections, with alternating 
signs, such that convergence is poor at any reasonable
temperature. Of course, at least without light dynamical fermions, 
the full pressure
can still be obtained with 4d finite temperature lattice
simulations~\cite{boyd}. However, in order to really understand the properties
of the QCD plasma phase, one would also like to have some
analytic understanding of the origin of this result.

A way of at least understanding why the convergence is poor
is the observation that when $\alpha_s=g^2/(4\pi) \ll 1$, 
the system undergoes dimensional 
reduction~\cite{dr,rold,hl,generic,bn,ad}, and its
static long wavelength ``soft'' or ``light''
degrees of freedom can be described by a
three-dimensional (3d) effective field theory, 
\[
{\cal L}_\rmi{3d} = % \int\!d^3\! x\,\biggl( 
\fr12 \tr F_{ij}^2
+ \tr [D_i,A_0]^2 + 
m_D^2\tr A_0^2 +\lambda_A(\tr A_0^2)^2, % \biggr),
\la{leff}
\]
where $m_D^2\sim g^2T^2, \lambda_A\sim g^4 T$ 
are parameters  computed perturbatively up to 
optimised next-to-leading order level~(see below).
This effective theory is confining, 
therefore non-perturbative~\cite{linde,gpy}.  
In~\cite{bn} ${\cal L}_\rmi{3d}$
was used to reproduce
the perturbative free energy up to order ${\cal O}(g^5T^4)$~\cite{az,zk}, 
and the bad convergence was shown to be due precisely to these 
degrees of freedom.

Our objective here is to study the free energy of QCD by including 
the dominant badly convergent contributions from
${\cal L}_\rmi{3d}$ non-perturba\-tively, to all orders, by using 
lattice Monte Carlo simulations. In this way, we can
find out how important the combined effect of the badly convergent 
series really is in the free energy.  

It is important to keep in mind that infrared sensitive effects
can be different in various quantities. For instance, the free
energy is dominated by ultraviolet degrees of freedom, 
and the long-distance effects we study here 
may turn out to be subdominant. Thus it would
be wrong to conclude that any approach 
which manages to reproduce the numerical data for the free energy
in a satisfactory way, would also reproduce other quantities. 
A good testing ground for this  
are the longest static correlation lengths in the QCD plasma:
they are fully non-perturbative, but it is already known that
the results of 4d simulations~\cite{dg} 
are reproduced precisely by the infrared
degrees of freedom we employ in ${\cal L}_\rmi{3d}$~\cite{rold,ad,mu}.

The relation of our approach to the other recent approaches
for the determination of the free energy 
of QCD~\cite{braaten,blaizot,peshier} can be described as follows. 
At present, these approaches do not reproduce the known ${\cal O}(g^5T^4)$
result in the limit of a weak coupling, nor do they 
account for any genuine non-perturbative contributions. 
Thus large infrared effects are suppressed 
without an a priori justification; the justification comes
a posteriori through the reasonable agreement with numerical data. 
Our results here attempt at providing a theoretical 
understanding for why the long-distance 
contributions need not be important in the QCD pressure.  

\pagebreak

{\bf Method.}
The pressure or the free energy density of QCD is a quantity 
which formally gets contributions both from short-distance physics 
($l \lsim (\pi T)^{-1}$) and long-distance physics
($l \gsim (g T)^{-1}$). The separation of the free energy 
into these two different types of contributions has been 
discussed in detail in~\cite{bn}. Interactions between 
the short and long-distance modes account for the parameters 
of the effective long-distance theory ${\cal L}_\rmi{3d}$, and in addition 
there is an additive part coming directly from 
the short-distance modes, as we will presently specify. 

To describe the effects of the short-distance modes 
in detail, we find it useful to introduce the 
dimensionless parameters $y=m_D^2/g_3^4, x=\lambda_A/g_3^2$, 
where $g_3^2$ is the gauge coupling within the effective theory.
In terms of the physical parameters $T,\lambdamsbar$ of QCD, 
next-to-leading order ``fastest apparent convergence''
optimised perturbation theory tells that~\cite{ad} 
(for a number of flavours $N_f=0$ and colours $N_c=3$),
\ba
\frac{g_3^2}{T} & = &  \frac{8\pi^2}{11\ln(6.742 T/\lambdamsbar)}, 
\la{params} \\
x & = &  \frac{3}{11\ln(5.371 T/\lambdamsbar)},
\quad y  =  
\frac{3}{8\pi^2 x}+\frac{9}{16\pi^2}. \la{ydr}
\ea
%For the mass parameter $y$ which has a part depending on the 
%regularization scheme, we have assumed the $\msbar$ scheme 
%with the scale parameter $\bmu_\rmi{3d} = g_3^2$. 

The result of~\cite{bn}, \eq(36),  
can now be expressed as follows. Using
the $\msbar$ scheme with the scale parameter $\bmu_\rmi{3d}$, 
let us compute the dimensionless quantity
\be
{\cal F}_{\rmi{$\msbar$}}(x,y) = 
-\frac{1}{Vg_3^6} \ln \biggl[ \int \! {\cal D}A 
\exp(-\int d^3x {\cal L}_\rmi{3d})\biggr], 
\ee
where $V$ is the volume. 
The pressure can then be expressed as
(we have here again put $N_f=0, N_c=3$) 
\ba
p(T) & = & p_0(T) \times \biggl[ 1 -\fr52 x  \nn
& - & \fr{45}{8\pi^2} 
\Bigl(\fr{g_3^2}T\Bigr)^3 
\Bigl( {\cal F}_{\rmi{$\msbar$}}(x,y) -24\fr{y}{(4\pi)^2} 
\ln\fr{\bmu_\rmi{3d}}T \Bigr) \biggr] \;,
\la{pressure}
\ea
where $p_0(T) = (\pi^2T^4 /45)(N_c^2-1+(7/4) N_c N_f)$
is the non-interacting Stefan-Boltzmann result.
The $\bmu_\rmi{3d}$-dependence here is cancelled by that in
${\cal F}_{\rmi{$\msbar$}}(x,y)$.

A few comments on \eq\nr{pressure} are in order. First of all, 
the term proportional to $y$ could also be written as 
$\sim {\cal O}(x^2)$, and at the present level of accuracy there
is no unique way of making a distinction. We have chosen
the present form
because the relatively large logarithmic term is then
dealt with in connection with ${\cal F}_{\rmi{$\msbar$}}$, 
whereby cancellations occur. Second, 
strictly speaking $\ln({\bmu_\rmi{3d}}/T)$
should be replaced with $\ln({\bmu_\rmi{3d}}/T)+\delta$, but
$\delta = 
\gamma_E - \ln 2 -{41}/{2160}  -({17}/{72})\ln 2\pi - 
({37}/{36}) [\ln\zeta]'(2)+ 
({19}/{72}) [\ln\zeta]'(4) \approx 1.35\times 10^{-4}$
can be ignored for all practical purposes. Finally, 
with the expressions available at present, the relation 
in \eq\nr{pressure} has an error starting at order ${\cal O}(g^6)$, 
corresponding to ${\cal O}(1/(4\pi)^4)$ within the parentheses. 
This correction is however from short-distance physics alone, 
and we shall ignore it here. 

By \eqs\nr{params}, \nr{ydr}, \nr{pressure}, the
perturbative short-distance contribution to the
pressure has been accounted for
to a satisfactory level, and we are left with
evaluating the long-distance part, ${\cal F}_{\rmi{$\msbar$}}(x,y)$.
The perturbative expression for ${\cal F}_{\rmi{$\msbar$}}(x,y)$ is 
known up to 3-loop level, corresponding to 
${\cal O}(g^5T^4)$ accuracy in $p(T)$.  
Adding terms involving the scalar self-interaction $x$ to 
the result of~\cite{bn}, we can write 
\ba
 & & \frac{{\cal F}_{\rmi{$\msbar$}}(x,y)}{d_A} =   
\frac{y^{\fr32}}{4\pi}  
\biggl[-\frac{1}{3}\biggr] \nn
 & & + \frac{y}{(4\pi)^2}
\biggl[C_A\Bigl(\fr34 - \fr12 \ln 4y + \ln\frac{\bmu_\rmi{3d}}{g_3^2}\Bigr)
+ \frac{d_A+2}{4} x \biggr] \nn
 & & +  \frac{y^{\fr12}}{(4\pi)^3} 
\biggl[C_A^2\Bigl(\frac{89}{24}-\frac{11}{6}\ln 2+\frac{\pi^2}{6}\Bigr)
- C_A \frac{d_A+2}{2} \Bigl(\fr12 - \ln 4y \Bigr) x \nn 
& & + 
\frac{d_A+2}{2} \Bigl(\frac{10-d_A}{4} - \ln 16y \Bigr) x^2 \biggr]
+ \frac{\Delta {\cal F}_{\rmi{$\msbar$}}(x,y)}{d_A},
\la{Fpert}
\ea
where 
$d_A=N_c^2-1, C_A=N_c$, and $\Delta {\cal F}_{\rmi{$\msbar$}}(x,y)$
accounts for the higher order corrections. 
In terms of the 4d coupling constant, all 
contributions involving $x$ in \eq\nr{Fpert} are
of order ${\cal O}(g^6)$ or higher, while the terms
$\sim y^{\fr32}, y\ln y, y^{\fr12}$ are 
of orders $g^3, g^4\ln (1/g), g^5$, respectively.  

As is well known~\cite{az,zk,bn}, the convergence of the perturbative 
expansion in \eq\nr{Fpert} is quite poor when values of $x,y$ corresponding
to any reasonable physical 
temperature $T/\lambdamsbar$ are chosen. For future reference, 
we illustrate this in \fig\ref{fig:pert}. We have used 
\eqs\nr{params}, \nr{ydr}, \nr{pressure} together with terms up to 
order $y^{\fr12}$ in \eq\nr{Fpert}. 

%%%%%%%%%%%%%%%%%%%%%%%%%%%%%%%%% FIGURE
\begin{figure}[t]

%\vspace*{-1.5cm}

\centerline{\epsfxsize=6cm%\hspace*{-1cm}
\epsfbox{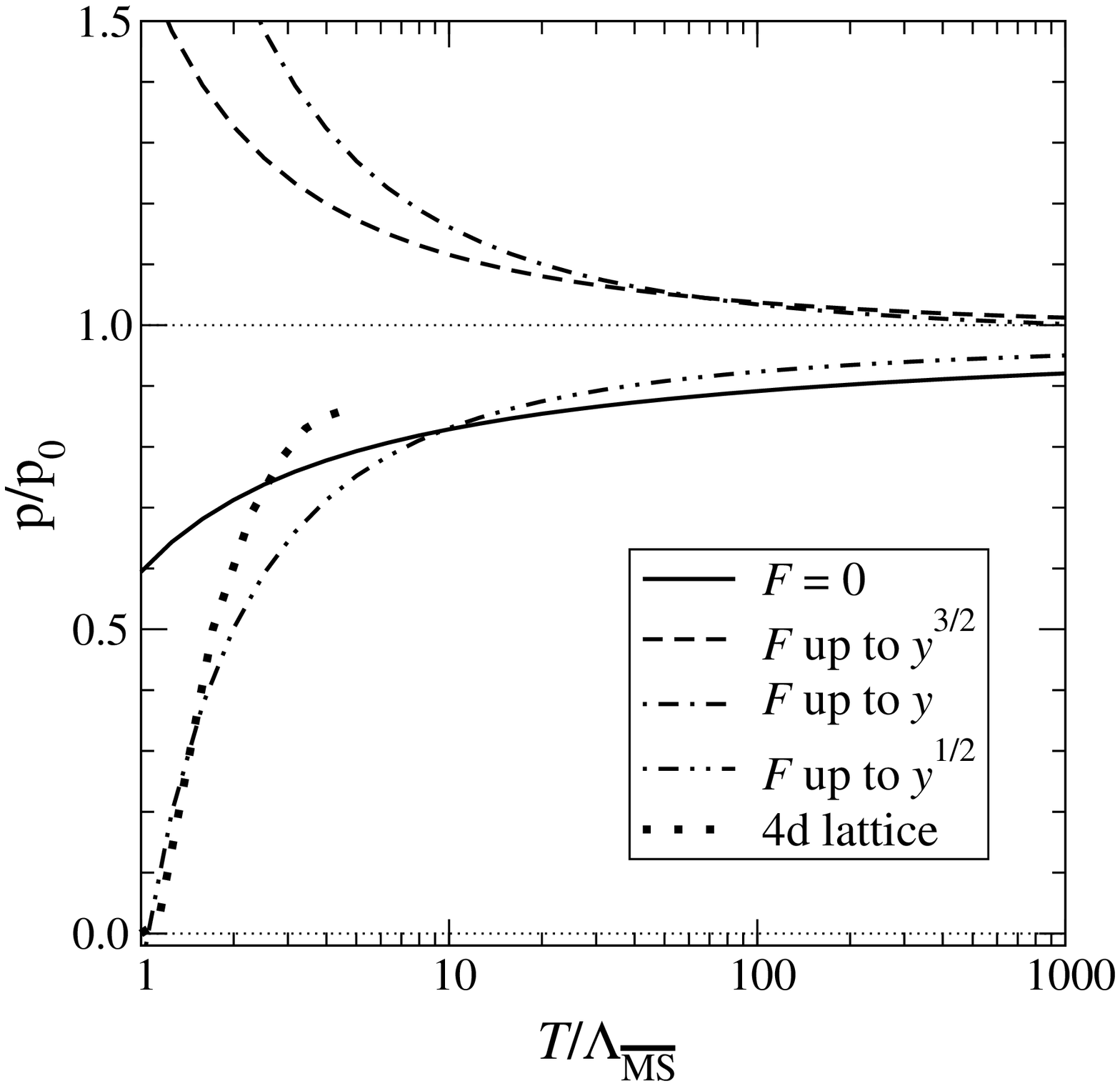}}

%\vspace*{-5cm}

\caption[a]{The pressure in \eq\nr{pressure}, with the 
long-distance part from \eq\nr{Fpert} included to 
various loop orders. The 4d lattice results are from the 
first reference in~\cite{boyd}. It should be noted that they have 
a normalisation ambiguity at low temperatures $T\lsim T_c$ allowing
for a small shift of the curve.}
%According to the 3rd Ref.\ in~\cite{boyd}, $T_c\lambdamsbar=1.14(4)$.}  
\la{fig:pert}
\end{figure}
%%%%%%%%%%%%%%%%%%%%%%%%%%%%%%%%%%%%

The idea of our approach of improving the determination 
of ${\cal F}_{\rmi{$\msbar$}}(x,y)$ is the following. 
We write 
\ba
\Delta {\cal F}_{\rmi{$\msbar$}}(x,y) & = &  
\Delta {\cal F}_{\rmi{$\msbar$}}(x_0,y_0) \nn 
& & + 
\int_{y_0}^{y} \! dy\, \biggl( 
\frac{\partial \Delta {\cal F}_{\rmi{$\msbar$}}}{\partial y} + \frac{d x}{dy} 
\frac{\partial \Delta {\cal F}_{\rmi{$\msbar$}}}{\partial x}
\biggr), \la{Ffull}
\ea
where $y=y(x)$ is defined in \eq\nr{ydr}. The partial derivatives 
are now given by adjoint Higgs field condensates:
\be
\frac{\partial \Delta {\cal F}_{\rmi{$\msbar$}}}{\partial y} = 
\biggl\langle \frac{\tr A_0^2}{g_3^2} \biggr\rangle_{\rmi{$\msbar$}}-
\biggl\langle \frac{\tr A_0^2}{g_3^2} \biggr\rangle_{\rmi{$\msbar$, pert}},
\la{A0}
\ee
where 
$\langle \tr A_0^2 /g_3^2 \rangle_{\rmi{$\msbar$, pert}}$
is the perturbative result up to ${\cal O }(y^{-\fr12})$, 
obtained by taking a derivative of \eq\nr{Fpert} with 
respect to $y$. In the case of 
$\partial \Delta {\cal F}_{\rmi{$\msbar$}}/\partial x$, 
a similar relation is obtained but with the condensate 
$\langle (\tr A_0^2)^2 \rangle$.

On the other hand, with a computation in lattice perturbation theory, 
a condensate measured in lattice Monte Carlo simulations can be related
to the condensates $\langle \tr A_0^2 \rangle_{\rmi{$\msbar$}}$, 
$\langle (\tr A_0^2)^2 \rangle_{\rmi{$\msbar$}}$. 
Due to the super-renormalizable nature
of ${\cal L}_\rmi{3d}$, such analytic 
relations can be computed exactly near the 
continuum limit~\cite{framework,moore_a}.

Thus, we need to evaluate the condensates on the lattice, 
transform the result to the $\msbar$ scheme, and perform finally
the integration in \eq\nr{Ffull} numerically. When added 
to $\Delta {\cal F}_{\rmi{$\msbar$}}(x_0,y_0)$, we obtain 
a non-perturbative result, which we can plug into \eq\nr{pressure}.

What remains is to determine the integration constant
$\Delta {\cal F}_{\rmi{$\msbar$}}(x_0,y_0)$. The idea is that
despite the bad convergence shown in ~\fig\ref{fig:pert}, 
at high enough temperatures the form of 
$\Delta {\cal F}_{\rmi{$\msbar$}}(x_0,y_0)$ is known. Indeed, 
inspecting the general structure of \eq\nr{Fpert}, 
we know that
\be
\Delta {\cal F}_{\rmi{$\msbar$}}(x_0,y_0) = 
\frac{e_0}{(4\pi)^4} d_A C_A^3 \Bigl(1 + 
{\cal O} (\frac{x_0}{C_A}, \frac{C_A}{4\pi {y_0}^{\fr12}})
\Bigr). \la{DFpert}
\ee
Here $e_0$, containing an unknown logarithmic dependence
on $y_0$, represents 
the famous non-perturbative ${\cal O}(g^6T^4)$ term~\cite{linde}. 
Suppose now that we
choose $T\equiv T_0\sim 10^{11} \lambdamsbar$, corresponding to 
$x_0=1.0\times 10^{-2}$, $y_0=3.86$. Then 
the higher order terms in \eq\nr{DFpert} are 
expected to be subdominant, since $C_A/(4\pi {y_0}^{\fr12})\sim 0.1$
and $x_0/C_A \sim 0.01$, and we only need to know $e_0$. 

The main error sources of this non-perturbative and unambiguous 
setup are as follows:

(a) Even though in principle 
an independent non-perturbative determination of $e_0$ 
is possible for instance
by measuring the condensate $\langle \tr F_{ij}^2 \rangle$ 
along the lines in~\cite{klpr}, doing this systematically 
requires a 4-loop computation in lattice perturbation theory, and this 
is beyond our scope here.
Therefore we will treat $e_0$ as a free integration
constant whose magnitude will be fixed below. 

(b) Due to the smallness of $x/C_A$, we will also ignore here
the term arising from $\partial \Delta {\cal F}_{\rmi{$\msbar$}}/\partial x$
in \eq\nr{Ffull}. %This is in any case of high perturbative order. 

(c) The numerical procedure introduces small statistical errors, as well as
systematic errors from the extrapolations to the infinite volume
and continuum limits. 

(d) Finally, we should of course remember that 
the effective theory ${\cal L}_\rmi{3d}$
loses its accuracy when higher order operators not included become 
important. In fact, for $N_f=0$ the QCD phase transition 
is related to the so called Z(3)-symmetry~\cite{gpy,ka1}, 
and this symmetry is not fully reproduced 
by ${\cal L}_\rmi{3d}$~\cite{ad,su3adj}, without all the higher order 
operators. There are many indications, however, that the
effective theory should be rather accurate down to 
low temperatures, $T\sim 2 T_c$~\cite{rold,ad,mu}. Below
that, some other effective description may apply (see, e.g.,~\cite{rdp}).

\vspace*{0.5cm}

{\bf Numerical results.} 
After this background, we show in \fig\ref{fig:a2values}
the difference in \eq\nr{A0}, measured with lattice 
simulations. This result is then used in \eq\nr{Ffull} to obtain 
$\Delta {\cal F}_{\rmi{$\msbar$}}(x,y)$. When added 
to \eqs\nr{Fpert}, \nr{pressure}, 
we obtain \fig\ref{fig:press3}. As discussed above, 
the boundary value at (almost) infinite temperature, determined
by $e_0$, is for the moment a free parameter. 

%%%%%%%%%%%%%%%%%%%%%%%%%%%%%%%%% FIGURE
\begin{figure}[t]

%\vspace*{-1.5cm}

\centerline{\epsfxsize=6cm%\hspace*{-1cm}
\epsfbox{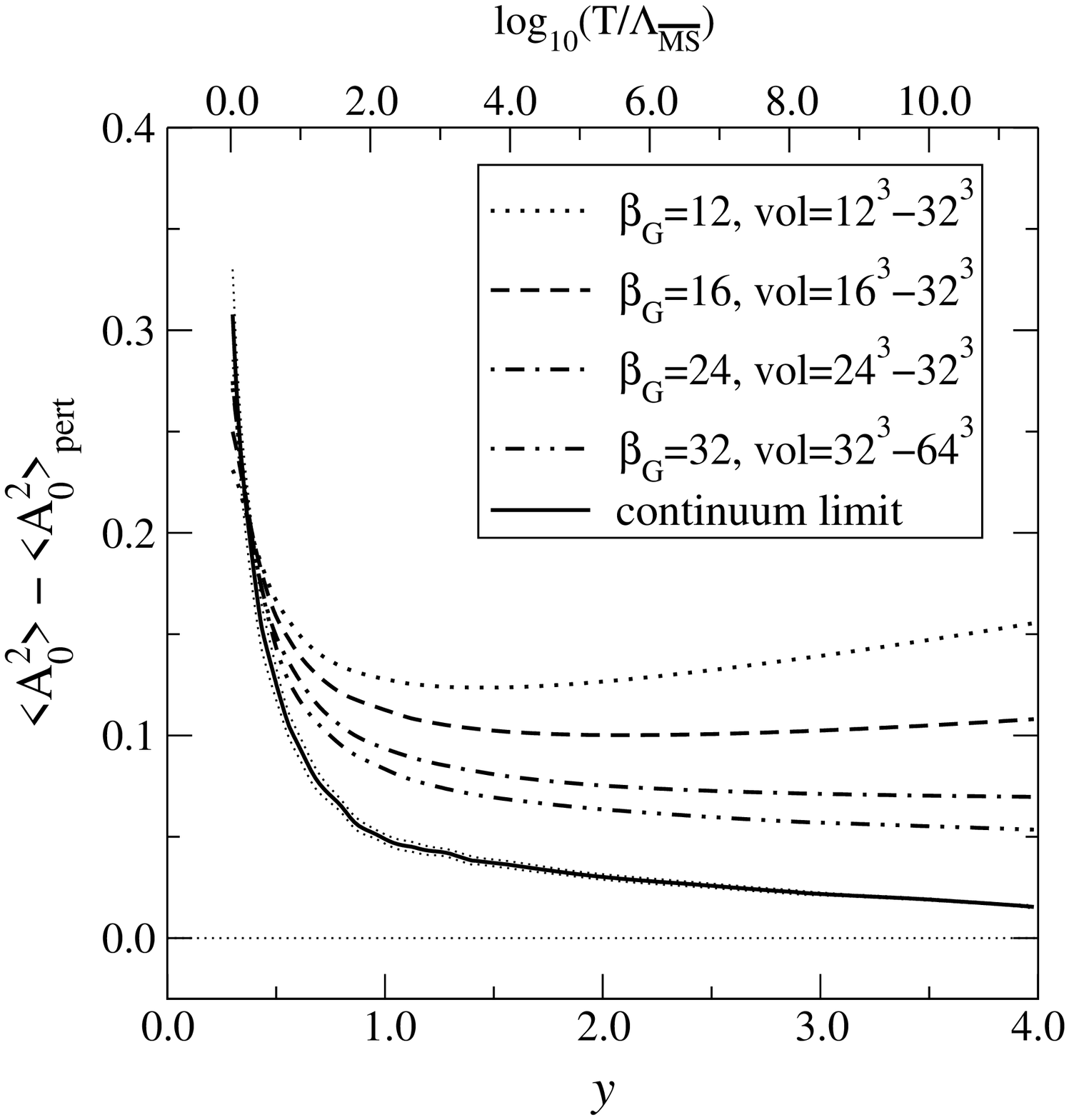}}

%\vspace*{-5cm}

\caption[a]{The difference in \eq\nr{A0}. 
Here $\beta_G=6/(g_3^2 a)$, where $a$ 
is the lattice spacing, and the continuum limit 
corresponds to the extrapolation $\beta_G\to\infty$.}
\la{fig:a2values}
\end{figure}
%%%%%%%%%%%%%%%%%%%%%%%%%%%%%%%%%%%%

%%%%%%%%%%%%%%%%%%%%%%%%%%%%%%%%% FIGURE
\begin{figure}[t]

%\vspace*{-1.5cm}

\centerline{\epsfxsize=6cm%\hspace*{-1cm}
\epsfbox{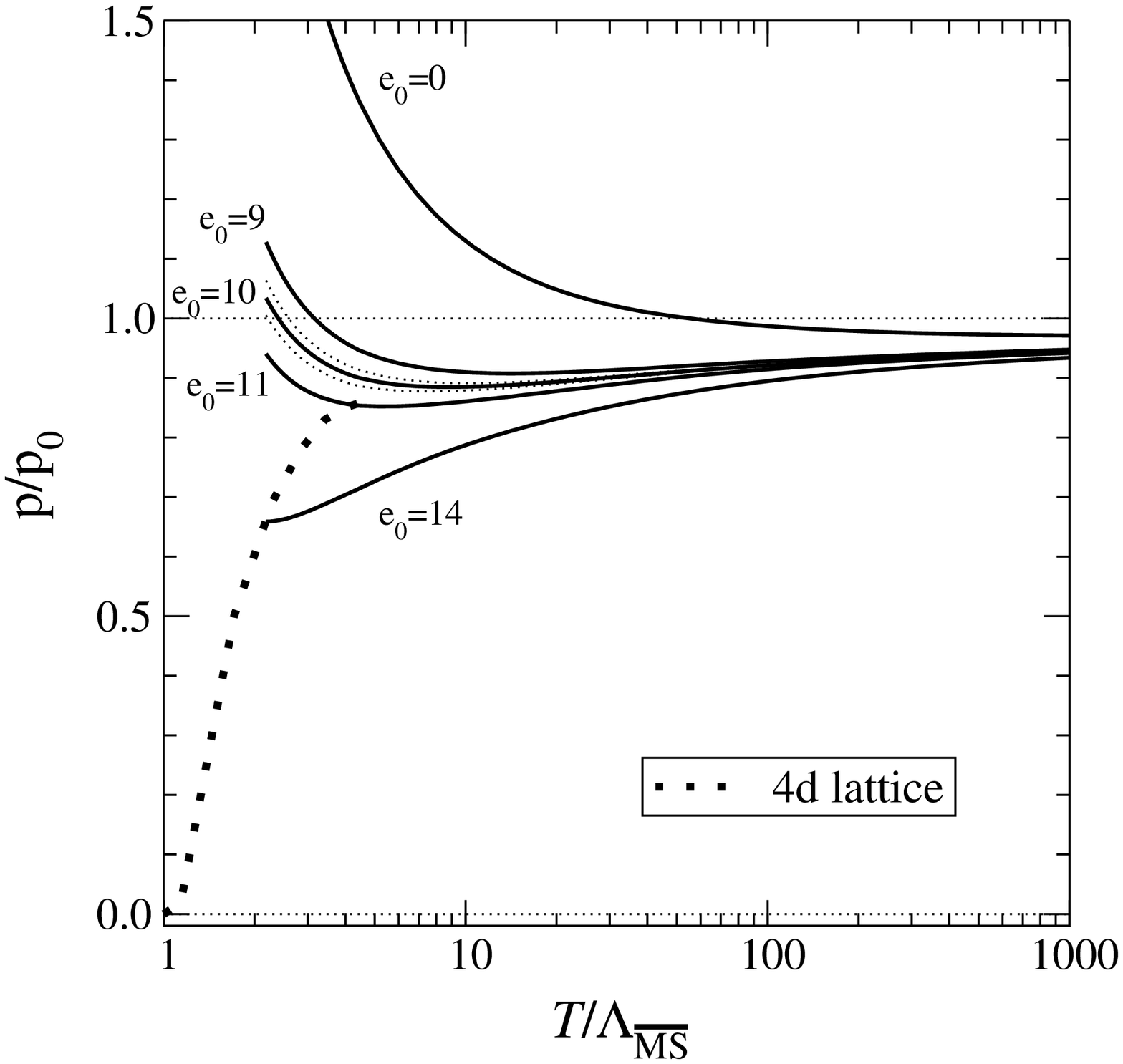}}

%\vspace*{-5cm}

\caption[a]{The pressure after the inclusion of 
$\Delta {\cal F}_{\rmi{$\msbar$}}(x,y)$ from \eq\nr{Ffull}.
Statistical errors are shown only for $e_0=10$.}  
\la{fig:press3}
\end{figure}
%%%%%%%%%%%%%%%%%%%%%%%%%%%%%%%%%%%%

We observe that at low temperatures the outcome depends strongly
on the value of $e_0$. The correct value would appear to be 
$e_0\approx 10.0\pm 2.0$. Even then, the present results 
lose their accuracy at $T \sim 5 T_c$, but seem to work 
well above this.
Exploiting the full power of the dimensionally
reduced theory down to its limit $T \sim 2T_c$ would 
also necessitate the inclusion of $\langle (\tr A_0^2)^2\rangle$. 

%\vspace*{0.5cm}

{\bf Discussion.}
In 4d lattice simulations, there is a
(numerically small) ambiguity in the determination
of the pressure, because only pressure differences can be measured, 
and thus an integration constant has to be specified at low temperatures
in a non-perturbative regime. Here we fix the integration constant by 
starting from the opposite direction, from very  high temperatures. 
This allows to determine all quantities in terms of $T/\lambdamsbar$ 
and the number of fermion flavours, without ambiguities. We can also
address a huge range of temperatures, unlike 4d simulations which
can only go up to $T \sim \mbox{a few} \times T_c$.

The result of our procedure is summarised by \eqs\nr{pressure}--\nr{A0}
and \fig\ref{fig:press3}. We draw two important conclusions. The first
is that the outcome depends strongly on the 
non-perturbative contribution of order ${\cal O}(g^6T^4)$~\cite{linde}, 
as can be observed from the $e_0$-dependence in \fig\ref{fig:press3}.
The value of $e_0$ could in principle be determined by a well-defined 
procedure, although in practice it is a project of considerable 
technical complication. But our present study provides an estimate 
for what the result should be. The order of magnitude ${\cal O}(10)$ 
seems reasonable, since it is known from other contexts such as the 
Debye mass~\cite{mu} that non-perturbative constants tend to be large.

The second is that when the large non-perturbative ${\cal O}(g^6T^4)$
term is summed together with the set of all higher order terms determined
via $\langle \tr A_0^2\rangle$, then these 
long-distance contributions 
almost cancel at $T \gsim 30\lambdamsbar$! Indeed, the sum, the 
curve with $e_0 \sim 10$ in \fig\ref{fig:press3}, does not differ there 
much from the term ${\cal O}(y^{\fr12})$ in \fig\ref{fig:pert}. For 
smaller temperatures, $5 \lambdamsbar \lsim T \lsim 30 \lambdamsbar$, 
on the other hand, only our numerical results are trustworthy. 

Finally, we also find that although the dependence on 
the effective scalar self-coupling $x$ is of high perturbative order, 
in practice it is expected to play a role as one approaches $T_c$. 
Its contribution can be obtained from 
the condensate $\langle (\tr A_0^2)^2 \rangle$. 
To relate this to the $\msbar$
scheme requires again a perturbative 4-loop computation.

Let us end with a philosophical note. When one wants to 
understand 4d simulation results, one could argue that
one should aim at almost fully analytical 
resummations~\cite{braaten,blaizot,peshier}. However, we suspect that these
are unavoidably specific for the particular observable 
considered: they may work for the entropy or pressure because the result is 
short-distance dominated, but would fail for instance for Debye
screening where long-distance effects are dominant. 
It seems to us that it may ultimately be more useful to obtain a unified 
understanding of the relevant degrees of freedom in the system, even if 
some observables have to be evaluated numerically. 

%%%%%%%%%%%%%%%%%%%%%%%%%%%%%%%%%%%%%%%%%%%%%%%%%%%%%%%

\vspace*{0.5cm}

{\bf Acknowledgements.}
This work was supported by the TMR network {\em Finite
  Temperature Phase Transitions in Particle Physics}, EU contract no.\ 
FMRX-CT97-0122. We thank the Center for Scientific Computing, Finland, 
for resources, and D.\ B\"odeker, E.\ Iancu for discussions.

\end{document}